\documentclass[aps,pra,showpacs,amsmath,amssymb,amsfonts,tightenlines,superscriptaddress,twocolumn,lengthcheck]{revtex4-1}
\usepackage{amsmath}
\usepackage{amsfonts}
\usepackage{graphicx}
\usepackage{epstopdf}
\graphicspath{{fig6/}}
\usepackage{epsfig}
\usepackage{color}
\usepackage{txfonts}
\usepackage[colorlinks,citecolor=blue]{hyperref}
\usepackage{textcomp}
\begin{document}
\title{Millionfold improvement in multivibration-feedback optomechanical refrigeration via auxiliary mechanical coupling}
\author{Rui Xu}
\affiliation{College of Physics and Electronic Engineering, Institute of Solid State Physics, Sichuan Normal University, Chengdu 610068, P. R. China}
	
\author{Deng-Gao Lai}
\email{Corresponding author: denggaolai@foxmail.com}
\affiliation{Theoretical Quantum Physics Laboratory, RIKEN Cluster for Pioneering Research, Wako-shi, Saitama 351-0198, Japan}
	
\author{Bang-Pin Hou}
\affiliation{College of Physics and Electronic Engineering, Institute of Solid State Physics, Sichuan Normal University, Chengdu 610068, P. R. China}	

\author{Adam Miranowicz}
\affiliation{Theoretical Quantum Physics Laboratory, RIKEN Cluster for Pioneering Research, Wako-shi, Saitama 351-0198, Japan}
\affiliation{Institute of Spintronics and Quantum Information, Faculty of Physics, Adam Mickiewicz University, 61-614 Pozna\'{n}, Poland}

\author{Franco Nori}
\affiliation{Theoretical Quantum Physics Laboratory, RIKEN Cluster for Pioneering Research, Wako-shi, Saitama 351-0198, Japan}
\affiliation{RIKEN Center for Quantum Computing (RQC), 2-1 Hirosawa, Wako-shi, Saitama 351-0198, Japan}
\affiliation{Physics Department, The University of Michigan, Ann Arbor, Michigan 48109-1040, USA}
	
\begin{abstract}
The simultaneous ground-state refrigeration of multiple vibrational modes is a prerequisite of observing significant quantum effects of multiple-vibration systems. Here we propose how to realize a giant amplification in the net-refrigeration rates based on cavity optomechanics, and to largely improve the cooling performance of multivibration modes beyond the resolved-sideband regime. By employing an auxiliary mechanical coupling (AMC) between two mechanical vibrations, the dark mode, which is induced by the coupling of these vibrational modes to a common optical mode and cuts off cooling channels, can be fully removed. We use fully analytical treatments for the effective mechanical susceptibilities  and net-cooling rates, and find that when the AMC is turned on, the amplification of the net-refrigeration rates  can be observed by more than six orders of magnitude. In particular, we reveal that the simultaneous ground-state cooling beyond the resolved-sideband regime arises from the introduced AMC, without which it vanishes. Our work paves a route for the quantum control of multiple vibrational modes in the bad-cavity regime.

\end{abstract}

\maketitle

\section{Introduction}

Cavity optomechanics~\cite{Bowen2015CRCpress,Kippenberg2008Science,Meystre2013AP,Aspelmeyer2014RMP} provides a promising platform to explore mechanical properties of quantum system via optical means and to manipulate cavity-field statistics by mechanically changing the boundary of a cavity~\cite{Rabl2011PRL,Nunnenkamp2011,Liao2012PRA,Liao2013PRA,Liao2013,Wang2013PRL,Liu2013PRL0,Vitali2007PRL,Agarwal2010PRA,Genes2011PRA,Cirio2017PRL,Xu12015PRA,Hou2015PRA,Zou2020OE,Bai2016PRA,Yang2017OE,Lai2020PRAOMIT,Xu2021PRA,Qian2021PRA,Wu2018PRApplied,Qin2018PRL,Zippilli2018PRA,Wang2015,Li2019,Jiao2020}.
As a prominent application closely related to this optomechanical platform, optomechanical refrigeration has become a hot research topic~\cite{Wilson-Rae2007PRL,Marquardt2007PRL,Genes2008PRA,Chan2011Nature,Teufel2011Nature,Clarkl2017Nature}. This is due to the fact that for observing significantly quantum effects of systems, a prerequisite is to cool these systems to their quantum ground states by effectively suppressing their thermal noise.
Until now, cooling a single mechanical mode to its quantum ground state of optomechanical systems, has been mainly achieved by the resolved-sideband-refrigeration~\cite{Wilson-Rae2007PRL,Marquardt2007PRL,Genes2008PRA,Chan2011Nature,Teufel2011Nature,Clarkl2017Nature} and feedback-aided-refrigeration~\cite{Mancini1998PRL,Steixner2005PRA,Bushev2006PRL,Rossi2017PRL,Rossi2018Nature,Conangla2019PRL,Tebbenjohanns2019PRL,Sommer2019PRL,Guo2019PRL,Sommer2020PRR} mechanisms, which are preferable in the good-cavity and bad-cavity regimes, respectively.

In recent years, much attention has been paid to multivibration-mode systems~\cite{Bowen2015CRCpress,Kippenberg2008Science,Meystre2013AP,Aspelmeyer2014RMP}. This is because these systems can provide an incomparable platform to investigate topological energy transfer~\cite{Xu2016Nature}, macroscopic mechanical coherence~\cite{Mancini2002PRL,Tian2004PRL,Massel2012Nc,Mari2013PRL,Matheny2014PRL,Zhang2015PRL,Ockeloen-Korppi2018,Riedinger2018Nature,Stefano2019PRL,Qin2019npj,Kotler2021Science,Ockeloen-Korppi2021Science}, and quantum many-body effects~\cite{Heinrich2011PRL,Xuereb2012PRL,Ludwig2013PRL,Xuereb2014PRL,Xuereb2015NJP,Mahmoodian2018PRL,Rabl2010NP}. In particular, they have been widely applied in high-performance sensors~\cite{Massel2011Nature,Huang2013PRL,Bernier2016PRL}, quantum mechanical computers~\cite{Masmanidis2007science,Yamaguchi2008NN}, and nonreciprocal devices~\cite{Malz2018PRL,Shen2016NP,Shen2018NC,Shen2022arXiv,Fang2017NP,Xu2019Nature,Mathew2018arXiv,Yang2020NC,SanavioPRB2020}.
These applications relevant to multivibration-mode systems, however, are fundamentally limited by thermal noise. To effectively suppress these thermal noise, simultaneously cooling these multivibration systems to their quantum ground states, becomes an important and urgent task. Although cooling a single vibrational mode to its quantum ground state has been realized in both optical~\cite{Chan2011Nature,Teufel2011Nature,Clarkl2017Nature,MXu2020PRL,Qiu2020PRL,Liu2013PRL,Liu2014PRA,Wang2011PRL,Li2011PRB,Yan2016PRA,Liao2011PRA,Machnes2012PRL,Lai2020PRARC,Lai2018PRA,Lai2021PRA1,Lai2021PRA2} and microwave~\cite{Grajcar2008PRB,Zhang2009PRA,Liberato2011PRA,Xue2007PRB,You2008PRL,Nori2008NP,Xiang2013RMP} domains, the simultaneous cooling of multiple vibrational modes remains an outstanding challenge in cavity optomechanics. The physical origin behind this challenge can be explained
by cooling suppression due to dark modes~\cite{Scully1997QO,Agarwal2013QO}, which are induced by the coupling of multiple vibrational modes to a common cavity-field mode~\cite{Shkarin2014PRL,Sommer2019PRL,Massel2012Nc,Genes2008NJP,Ockeloen-Korppi2019PRA,Kuzyk2017PRA,Huang2021arXiv,Liu2021arXiv,Lai2022PRL,Lai2022PRR1,Lai2022PRR2}. 

%Recently, based on the resolved-sideband-cooling mechanism, a dark-mode-breaking method using an auxiliary cavity has been developed for cooling multiple mechanical resonators in the good-cavity regime~\cite{}. However, the answer to the question of whether one can remove the dark mode by exploiting other methods and employ the feedback-cooling technique to simultaneously cool these mechanical resonators  in the bad-cavity regime is yet unclear.

\begin{figure*}[ht]
\centering
\includegraphics[width=17.5cm]{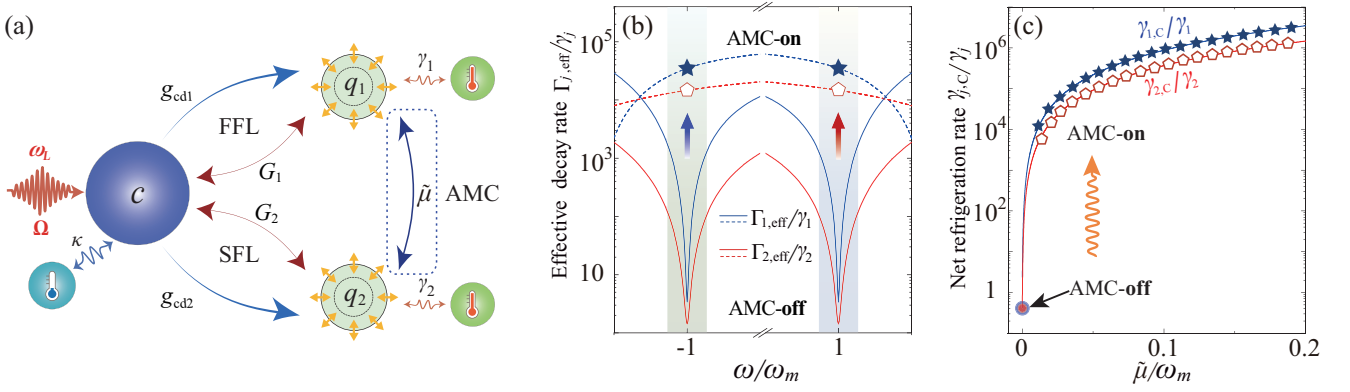}
\hfill
\caption{(a) Schematic diagram of a three-mode optomechanical system.  An optical mode $c$ (with frequency $\omega_{c}$ and decay rate $\kappa$)  is coupled to two vibrational modes $q_{j}$ (with frequency $\omega_{j}$ and damping rate $\gamma_{j}$) via radiation-pressure couplings (with strength $G_{j}$), and the two vibrational modes are coupled to each other through an AMC (with strength $\tilde{u}$). The cavity-field mode interacts with the vacuum bath, and the two vibrational modes are connected to their high-temperature heat baths. Note that the FFL denotes the strength of the first feedback loop, and the SFL describes the strength of the second feedback loop. (b) The effective mechanical damping $ \Gamma_{j,\mathrm{eff}}(\omega)$ of the $j$th vibrational mode versus the frequency $ \omega $, when the system operates  without  ($ \tilde{\mu}=0$, solid curves) and with  ($ \tilde{\mu}/\omega_{m}=0.02$, dashed curves) the AMC. (c) The net-refrigeration rate $ \gamma_{j,\mathrm{C}}(\omega)$ of the $j$th vibrational mode as a function of the AMC strength $ \tilde{\mu} $. The parameters used are:  $\omega_{j=1,2}=\omega_{m}=2\pi\times10^{7}$, $ G_{1}/\omega_{m}=0.4 $, $ G_{2}=0.7G_{1} $, $ g_{\mathrm{cd1}}=1 $, $ g_{\mathrm{cd2}}=0.6 $, $\kappa/\omega_{m}=4$, $\gamma_{j}/\omega_{m}=10^{-6}$, $\vartheta=0.8$, $\omega_{\mathrm{fb}}/\omega_{m}=3$, and $ \bar{n}_{1}=\bar{n}_{2}=10^{3}$.
 }\label{FIG1}
\end{figure*}

In this paper, based on the feedback-cooling mechanism, we proposed a dark-mode-removing method to achieve the simultaneous ground-state cooling of the two mechanical modes in the unresolved-sideband regime. This is realized by employing auxiliary mechanical coupling (AMC) to break the symmetry of the system, and, then, both dark and bright modes can be effectively controlled. By obtaining the exact analytical results of the net-cooling rates, effective mechanical susceptibilities, and steady-state mean phonon numbers,
we find that when the AMC is turned on in the system, a giant amplification in the net-refrigeration rates can be observed. Specifically, these net-refrigeration rates can be amplified to more than six orders of magnitude by properly tuning the AMC strength.

In particular, we show that the tremendously amplified net-refrigeration rates can result in a giant improvement for the refrigeration performance of the mechanical modes. Without AMC, the two vibrational modes cannot be efficiently cooled due to an inefficient net-cooling rate. However, when the AMC is turned on, the simultaneous ground-state cooling of these vibrations is achieved beyond the resolved-sideband regime, owing to the giant amplification in the net-cooling rates. Remarkably, we reveal that the larger is the feedback-loop strength of the resonator, the better is the cooling efficiency of this resonator.
Physically, the introduced AMC offers an effective strategy to remove the dark mode and, in turn, to rebuild cooling channels for extracting thermal phonons stored in the dark mode. This study could pave the way for studying quantum control and observing quantum mechanical coherence effects involving multiple vibrational modes.

\section{Model and hamiltonian}\label{II}

As shown in Fig.~\ref{FIG1}(a), we consider a three-mode optomechanical system, in which two vibrational modes are optomechanically coupled to a common optical mode. An AMC between the two vibrational modes is introduced to improve the net-cooling rates and the cooling performance of the system. To control the system, an external control laser with amplitude $\Omega$ and frequency $\omega_{L}$ is applied  to the optical cavity. The Hamiltonian of the system reads ($\hbar=1$)
\begin{align}
H_{0}=&\;\omega_{c}c^{\dag}c+\sum_{j=1}^{2}(\frac{1}{2m_{j}}p_{x,j}^{2}+\frac{1}{2}m_{j}\tilde{\omega}_{j}^{2}x_{j}^{2})
-\lambda_{1}c^{\dagger}cx_{1}-\lambda_{2}c^{\dagger}cx_{2}\notag\\&
+\mu( x_{1}-x_{2})^{2}+\Omega(c^{\dagger}e^{-i\omega_{L}t}+ce^{i\omega_{L}t} ), \label{HHH1}
\end{align}
where $c$ ($c^{\dag}$) denotes the annihilation (creation) operator of the optical mode. The operators $p_{x,j}$ ($j=1,2$) and $x_{j} $ are, respectively, the momentum and position operators of the $j$th vibrational mode, with  frequency $\tilde{\omega}_{j}$ and mass $m_{j}$. The $\lambda_{j}$ terms describe the optomechanical interactions between the optical mode and the $j$th vibrational mode, and the $\mu$ term denotes the AMC between the two vibrations. Note that the FFL denotes the strength of the first feedback loop, and the SFL describes the strength of the second feedback loop.

For the convenience of studying the system, we introduce the dimensionless position ($q_{j}=\sqrt{m_{j}\omega_{j}}x_{j}$) and  momentum  ($p_{j}=\sqrt{1/(m_{j}\omega_{j})}p_{x,j}$) operators, and the normalized resonance frequencies $\omega_{j}=\sqrt{\tilde{\omega}_{j}^{2}+2\mu/{m_{j}}}$ for $j=1,2$. In a rotating frame, defined by $\rm exp$$(-i\omega_{L}c^{\dag}ct)$, the system Hamiltonian~(\ref{HHH1}) becomes
\begin{align}
	H=&\;\Delta_{c}c^{\dagger}c+\sum_{j=1}^{2}\dfrac{\omega_{j}}{2}(q_{j}^{2}+p_{j}^{2})-\tilde{\lambda}_{1}c^{\dagger}cq_{1}-\tilde{\lambda}_{2}c^{\dagger}cq_{2}\notag\\&-2\tilde{\mu}q_{1}q_{2}+\Omega(c^{\dagger}+c),\label{HHH2}
\end{align}
where $\tilde{\lambda}_{1}=\lambda_{1}\sqrt{1/(m_{1}\omega_{1})}$, $\tilde{\lambda}_{2}=\lambda_{2}\sqrt{1/(m_{2}\omega_{2})}$, $\tilde{\mu}=\mu\sqrt{1/(m_{1}m_{2}\omega_{1}\omega_{2})}$, and $\Delta_{c}=\omega_{c}-\omega_{L}$.
We here need to emphasize that the Hamiltonian in Eq.~(\ref{HHH2}) is the starting point of our analysis and numerical simulations.

\section{Langevin equations and steady-state  mean phonon numbers}\label{III}

In this section, we obtain the Langevin equations of the system, analyze a cold-damping feedback-cooling scheme, and derive the  steady-state average phonon numbers.
	
\subsection{Langevin equations}
We consider the case where the two vibrational modes are subjected to quantum Brownian forces, and the optical mode interacts with  their vacuum baths. Then, the quantum Langevin equations can be used to describe the evolution of the system:
\begin{align}
\dot{c}&=-[\kappa+i(\Delta_{c}-\tilde{\lambda}_{1}q_{1}-\tilde{\lambda}_{2}q_{2})]c-i\Omega+\sqrt{2\kappa}c_{in},\notag\\
\dot{q}_{j}&=\omega_{j}p_{j},~~j=1,2\notag\\
\dot{p}_{1}&=-\omega_{1}q_{1}+2\tilde{\mu}q_{2}+\tilde{\lambda}_{1}c^{\dagger}c-\gamma_{1}p_{1}+\xi_{1},\notag\\
\dot{p}_{2}&=-\omega_{2}q_{2}+2\tilde{\mu}q_{1}+\tilde{\lambda}_{2}c^{\dagger}c-\gamma_{2}p_{2}+\xi_{2},\label{HHH3}	
\end{align}
where the operators $c_{in}$ and $\xi_{j}$ are, respectively, the input noise operator of the cavity-field mode and the Brownian noise operator resulting from the coupling of the corresponding vibrational modes to the thermal baths. These noise operators satisfy zero mean values and the following correlation functions:
\begin{align}
&\langle c_{in}(t)c_{in}^{\dagger}(t^{'})\rangle=\delta(t-t^{'}),\notag\\
&\langle c_{in}^{\dagger}(t)c_{in}(t^{'})\rangle=\langle c_{in}^{\dagger}(t^{'})c_{in}(t)\rangle=0, \notag\\
&\langle \xi_{j}(t)\xi_{j}(t^{'}) \rangle=\frac{1}{2\pi}\dfrac{\gamma_{j}}{\omega_{j}}\int\omega e^{-i\omega(t-t^{'})}\left[1+\coth(\dfrac{\hbar\omega}{2k_{\mathrm{B}}T_{j}})\right]d\omega,\label{HHH4}
\end{align}
where $ k_{\mathrm{B}} $ is the Boltzmann constant, and $T_{j}$ is the thermal bath temperature associated with the $j$th vibrational mode.

We assume that the cavity is strongly driven, and this allows us to linearize the dynamics of the system by writing each operator as the sum of their averages and fluctuations, i.e., $A=\langle A\rangle_{ss}+\delta A$ for $A\in\{c, c^{\dag}, q_{j}, p_{j}\}$. By neglecting higher-order fluctuation terms, the linearized quantum Langevin equations, which are described by the quadrature fluctuations $\delta X=(\delta c^{\dag}+\delta c)/\sqrt{2}$ and $\delta Y=i(\delta c^{\dagger}-\delta c)/\sqrt{2}$, are given by
\begin{align}
\delta\dot{X}=&-\kappa\delta X+\Delta \delta Y+\sqrt{2\kappa}X_{in},\notag\\
\delta\dot{Y}=&-\kappa\delta Y-\Delta\delta X+G_{1}\delta q_{1}+G_{2}\delta q_{2}+\sqrt{2\kappa}Y_{in},\notag\\
\delta\dot{q}_{j}=&\;\omega_{j}\delta p_{j},~~j=1,2,\notag\\
\delta\dot{p}_{1}=&-\omega_{1}\delta q_{1}+2\tilde{\mu}\delta q_{2}+G_{1}\delta X-\gamma_{1}\delta p_{1}+\xi_{1},\notag\\
\delta\dot{p}_{2}=&-\omega_{2}\delta q_{2}+2\tilde{\mu}\delta q_{1}+G_{2}\delta X-\gamma_{2}\delta p_{2}+\xi_{2},\label{HHH5}
\end{align}
where $X_{in}=(\delta c_{in}^{\dag}+\delta c_{in})/\sqrt{2} $ and $ Y_{in}=i(\delta c_{in}^{\dagger}-\delta c_{in})/\sqrt{2}$ are the corresponding Hermitian input noise quadratures, and the parameter $\Delta=\Delta_{c}-\tilde{\lambda}\langle q_{1}\rangle_{ss}-\tilde{\lambda}\langle q_{2}\rangle_{ss}$ is the normalized effective driving detuning. Moreover, $ G_{1}=\sqrt{2}\tilde{\lambda}_{1}\langle c\rangle_{ss}$ and $ G_{2}=\sqrt{2}\tilde{\lambda}_{2}\langle c\rangle_{ss}$ are the effective optomechanical coupling strengths, with $\langle c\rangle_{ss}=-i\Omega/(\kappa+i\Delta)$. Note that the phase reference of the cavity field $\langle c\rangle_{ss}$ is assumed to be real and positive.

\subsection{Cold-damping feedback}

To realize the cold-damping feedback refrigeration, the case of $\Delta=0$ is considered, so that the highest sensitivity for the position measurements of the vibrational modes can be achieved~\cite{Mancini1998PRL,Steixner2005PRA,Bushev2006PRL,Rossi2017PRL,Rossi2018Nature,Conangla2019PRL,Tebbenjohanns2019PRL,Sommer2019PRL,Guo2019PRL,Sommer2020PRR}. This feedback-refrigeration mechanism is essentially different from the sideband-cooling mechanism requiring the red-sideband resonance, i.e., $\Delta=\omega_{j}$~\cite{Wilson-Rae2007PRL,Marquardt2007PRL,Genes2008PRA,Chan2011Nature,Teufel2011Nature,Clarkl2017Nature}. By using a negative-derivative feedback, the effective decay rate of the mechanical mode can be largely developed by the cold-damping feedback technique.

Physically, the position of the two mechanical modes is measured through the phase-sensitive detection of the cavity output field, and, then, the readout of the cavity output field is fed back onto the two vibrational modes by applying feedback forces. The intensity of these feedback forces is proportional to the time derivative of the output signal, and, therefore, to the velocity of the mechanical modes. Then, the linearized quantum Langevin equations in Eq. (\ref{HHH5}) become
\begin{align}
&\delta\dot{X}=-\kappa\delta X+\sqrt{2\kappa}X_{\mathrm{in}},\notag\\
&\delta\dot{Y}=-\kappa\delta Y+G_{1}\delta q_{1}+G_{2}\delta q_{2}+\sqrt{2\kappa}Y_{\mathrm{in}},\notag\\
&\delta\dot{q}_{j}=\omega_{j}\delta p_{j},\notag\\
\delta\dot{p}_{1}=&-\omega_{1}\delta q_{1}+2\tilde{\mu}\delta q_{2}+G_{1}\delta X-\gamma_{1}\delta p_{1}+\xi_{1}\notag\\&
-\int^{t}_{-\infty}g_{1}(t-s)\delta Y^{\mathrm{est}}(s)ds,\notag\\
\delta\dot{p}_{2}=&-\omega_{2}\delta q_{2}+2\tilde{\mu}\delta q_{1}+G_{2}\delta X-\gamma_{2}\delta p_{2}+\xi_{2}\notag\\&
-\int^{t}_{-\infty}g_{2}(t-s)\delta Y^{\mathrm{est}}(s)ds,\label{HHH6}	
\end{align}
where the convolution term $ \int^{t}_{-\infty}g_{j}(t-s)\delta Y^{\mathrm{est}}(s)ds$~$(j=1,2)$ denotes the feedback force acting on the $j$th vibrational mode. These feedback forces depend on the past dynamics of the detected quadrature $\delta Y$, which is driven by a  weighted sum of the fluctuations of the vibrational modes. Here, the causal kernel is defined by
\begin{align}
&g_{j}(t)=g_{\mathrm{cd},j}\dfrac{d}{dt}[\theta(t)\omega_{\mathrm{fb}}e^{-\omega_{\mathrm{fb}}t}],\label{HHH7}	
\end{align}
where $ g_{\mathrm{cd},j} $ and $ \omega_{\mathrm{fb}}$ are the dimensionless feedback gain and feedback bandwidth associated the $j$th vibrational mode, respectively. In Eq.~(\ref{HHH7}), we have assumed that the electronic loop can provide an instantaneous feedback onto the system, and this assumption is included in the argument of the Heaviside function $\theta(t)$~\cite{Genes2008PRA,Sommer2019PRL,Sommer2020PRR}. This assumes fast electronics that can respond much quicker than the oscillation time of the system~\cite{Genes2008PRA,Sommer2019PRL,Sommer2020PRR}. The estimated intracavity phase quadrature $ \delta Y^{\mathrm{est}} $ results from the homodyne measurement of the output quadrature $ \delta Y^{\mathrm{out}}(t) $.
Here we generalize the usual input-output relation,
\begin{align}
\delta Y^{\text{out}}(t)=\sqrt{2\kappa}\delta Y(t)-Y^{\text{in}}(t),
\end{align}
to the case of a nonunit detection efficiency by modeling a detector with quantum efficiency $\vartheta$ with an ideal detector preceded by a beam splitter (with transmissivity $\sqrt{\vartheta}$), which mixes the incident field with the uncorrelated vacuum field $Y^{\upsilon}(t)$. Then, we obtain the generalized input-output
relation,
\begin{align}
Y^{\text{out}}(t)=\sqrt{\vartheta}[\sqrt{2\kappa}\delta Y(t)-Y^{\mathrm{in}}(t)]-\sqrt{1-\vartheta}Y^{\upsilon}(t).
\end{align}
The estimated phase quadratures $\delta Y^{\text{est}}(t)$ are obtained as\begin{align}
&\delta Y^{\text{est}}(t)=\frac{Y^{\text{out}}(t)}{\sqrt{2\vartheta \kappa}}=\delta Y(t)-\frac{Y^{\text{in}}(t)+\sqrt{\vartheta^{-1}-1}Y^{\upsilon}(t)}{\sqrt{2\kappa}}.
\end{align}

Below, we seek the steady-state solution of Eq.~(\ref{HHH6}) by solving it in the frequency domain with the Fourier transform.  $r(t)=(1/2\pi)^{1/2}\int_{-\infty }^{\infty }e^{-i\omega t}\tilde{r}(\omega) d\omega$ (for $r=\delta X$, $\delta Y$, $\delta q_{j}$, $\delta p_{j}$, $\xi_{j}$, $X_{\text{in}}$, and $Y_{\text{in}}$), and consequently the quantum Langevin equations in Eq.~(\ref{HHH6}), with the cold-damping feedback, can be solved in the frequency domain. Based on the steady-state solution, we can calculate the spectra of the position and momentum operators for two mechanical modes, and, then, the steady-state mean phonon numbers in these resonators can be obtained by integrating the corresponding fluctuation spectra.

\subsection{Final average phonon numbers}

The final steady-state average phonon numbers in the $j$th vibrational mode can be obtained by
\begin{align}
&n^{f}_{j}=\dfrac{1}{2}[\langle \delta q^{2}_{j}\rangle+\langle \delta p^{2}_{j}\rangle-1],~~~ j=1,2, \label{HHH9}
\end{align}
where $\langle \delta q^{2}_{j}\rangle$ and $\langle \delta p^{2}_{j}\rangle$ are the variances of the position and momentum operators, respectively. We solve Eq.~(\ref{HHH6}) in the frequency domain and integrate the corresponding fluctuation spectra, and then, the corresponding variances can be obtained as
\begin{align}
&\langle \delta q^{2}_{j}\rangle=\dfrac{1}{2\pi}\int^{\infty}_{-\infty}S_{q_{j}}(\omega)d\omega,\notag\\
&\langle \delta p^{2}_{j}\rangle=\dfrac{1}{2\pi\omega^{2}_{j}}\int^{\infty}_{-\infty}\omega^{2}S_{q_{j}}(\omega)d\omega. \label{HHH10}
\end{align}
The fluctuation spectra of the position and momentum operators are defined by
\begin{align}
&S_{A}(\omega)=\int^{\infty}_{-\infty}e^{-i\omega\tau}\langle \delta A(t+\tau)\delta A(t)\rangle_{\mathrm{ss}}d\tau,~(A=q_{j},p_{j})\label{HHH11}
\end{align}
where $ \langle \cdot\rangle_{\mathrm{ss}} $ denotes the steady-state average of the system. The fluctuation spectra in the frequency domain are expressed  as
\begin{align}
&\langle \delta\tilde{A}(\omega)\delta\tilde{A}(\omega^{\textquotesingle})\rangle_{\mathrm{ss}}=S_{A}(\omega)\delta(\omega+\omega^{'}).\label{HHH12}
\end{align}
Thus, in the frequency domain, we can solve this system and obtain the analytical results of the steady-state average phonon numbers. 

\subsection{Dark-mode effect and its removing}

%For studying quantum cooling of the two mechanical resonators, the beam-splitting-type interactions (i.e., the rotating-wave interaction term) between these bosonic modes are expected to dominate the linearized couplings in this system, and hence we can simplify the Hamiltonian of the system by making the rotating-wave approximation (RWA). The linearized optomechanical Hamiltonian in the RWA takes the following form (discarding the noise terms)

We next show the dark-mode effect and its removing from the two-vibrational-mode optomechanical system. For convenience, we introduce the annihilation (creation) operators of the two vibrational modes: $b_{j}=(q_{j}+ip_{j})/\sqrt{2}$ [$b_{j}^{\dagger}=(q_{j}-ip_{j})/\sqrt{2}$]. In the process of optomechanical cooling, the beam-splitting-type interactions (corresponding to the rotating-wave interaction term) between these bosonic modes dominate the linearized couplings in this system. By considering a red-detuned driving of the cavity and performing the rotating-wave approximation (RWA), the Hamiltonian of the system can be simplified as (discarding the noise terms),
\begin{eqnarray}
H_{\text{RWA}}&=&\Delta \delta c^{\dagger}\delta c+\sum^{2}_{j=1}[\omega_{j}\delta b_{j}^{\dagger}\delta b_{j}+G_{j}(\delta c\delta b_{j}^{\dagger}+\delta b_{j}\delta c^{\dagger})] \nonumber\\
&&+\eta(\delta b_{1}^{\dagger}\delta b_{2}+\delta b_{2}^{\dagger}\delta b_{1}).\label{RWAH}
\end{eqnarray}

To show the dark-mode effect and its removing in the system, we discuss in detail the physical system when the AMC is absent ($\eta=0$) and present ($\eta\neq 0$), respectively

(i) To study the dark-mode effect, we first assume that the AMC is turned off (i.e., $\eta=0$). In this case, the system can induce a bright mode and a dark mode:
\begin{subequations}
\begin{align}
B_{+}=&\;(G_{1}\delta b_{1}+G_{2}\delta b_{2})/G_{0}, ~~~bright\\
B_{-}=&\;(G_{2}\delta b_{1}-G_{1}\delta b_{2})/G_{0},~~~dark\label{BDmodedef}
\end{align}
\end{subequations}
where $G_{0}=\sqrt{G^{2}_{1}+G^{2}_{2}}$. Then, the Hamiltonian in Eq.~(\ref{RWAH}) can be rewritten with the bright and dark modes as
\begin{eqnarray}
H_{\text{hyb}} &=&\Delta\delta c^{\dagger }\delta c+\sum_{j=\pm}\omega_{\pm}B_{\pm}^{\dagger}B_{\pm}+G_{+}
(\delta cB_{+}^{\dagger}+B_{+}\delta c^{\dagger })\nonumber\\
&&+G_{-}(B_{+}^{\dagger}B_{-}+B_{-}^{\dagger}B_{+}),\label{Parity}
\end{eqnarray}
where we introduced the resonance frequencies $\omega_{+} =(G^{2}_{1}\omega_{1}+G^{2}_{2}\omega_{2})/G_{0}^{2}$ and $\omega_{-} =(G^{2}_{2}\omega_{1}+G^{2}_{1}\omega_{2})/G_{0}^{2}$. The coupling strengths $G_{+}$ and $G_{-}$ are, respectively, defined as 
\begin{align}
G_{+}=\sqrt{G^{2}_{1}+G^{2}_{2}},~~~~G_{-}=G_{1}G_{2}(\omega_{1}-\omega_{2})/G_{0}^{2}.\label{ParityG}
\end{align}
We see from Eq.~(\ref{ParityG}) that when $\omega_{1}=\omega_{2}$, the mode $B_{-}$ is decoupled from the system and it becomes a dark mode.

(ii) We then turn on the AMC (i.e., $\eta\neq 0$), so that the dark-mode effect can be efficiently removed. We introduce the two new bosonic modes $\tilde{B}_{\pm}$ associated with the AMC, defined by
\begin{align}
\delta b_{1}=f\tilde{B}_{+}+h\tilde{B}_{-},~~~~\delta b_{2}=-h\tilde{B}_{+}+f\tilde{B}_{-},
\end{align}
and, then, the Hamiltonian in Eq.~(\ref{RWAH}) becomes
\begin{eqnarray}
H_{\text{RWA}}&=&\Delta \delta c^{\dagger }\delta c+\sum_{j=\pm}[\tilde{\omega}_{j}\tilde{B}_{j}^{\dagger}\tilde{B}_{j}+(\tilde{G}_{j}^{\ast}\delta c\tilde{B}_{j}^{\dagger}+\tilde{G}_{j}\tilde{B}_{j}\delta c^{\dagger})],\label{DigH}
\end{eqnarray}
where the resonance frequencies $\tilde{\omega}_{\pm}=(\omega _{1}+\omega _{2}\pm \sqrt{(\omega_{1}-\omega_{2})^{2}+4\eta ^{2}})/2$, and the redefined-coupling strengths $\tilde{G}_{\pm}$ are:
\begin{eqnarray}
\tilde{G}_{+}=fG_{1}-hG_{2},~~~~\tilde{G}_{-}=hG_{1}+fG_{2},\label{DigG}
\end{eqnarray}
with
$f =\frac{\vert\tilde{\omega}_{-}-\omega_{1}\vert}{\sqrt{(\tilde{\omega}_{-}-\omega_{1})^{2}+\eta^{2}}}$,
$h =\frac{\eta f}{\tilde{\omega}_{-}-\omega_{1}}$.
When $\omega_{1}=\omega_{2}=\omega_{m}$, the coupling strengths in Eq.~(\ref{DigG}) can be simplified as
\begin{eqnarray}
\tilde{G}_{\pm}=(G_{2}\pm G_{1})/\sqrt{2},\label{DDigGequwG12}
\end{eqnarray}
It is shown in Eqs.~(\ref{DigH}) and~(\ref{DDigGequwG12}) that the dark mode $\tilde{B}_{-}$ can be fully removed when the strengths of the two optomechanical coupling are different (i.e., $G_{1}\neq G_{2}$).
The underlying physical mechanism behind our proposed method can be explained as follows: By tuning the coupling strength between the optical mode and each mechanical mode, the symmetry of the system is broken and both bright and dark mechanical modes can be effectively manipulated.

\section{Cooling of two mechanical modes}\label{IV}

In this section,  we derive the analytical expressions of the effective mechanical susceptibilities and net-refrigeration rates, and study the cooling performance of the two vibrational modes.

\subsection{Analytical results of effective susceptibilities, cooling rates, and noise spectra}
We obtain the position fluctuation spectra of the two vibrational modes as
\begin{align}
S_{q_{j}}(\omega)=&\mid \chi_{j,\mathrm{eff}}(\omega)\mid^{2}[S_{\mathrm{fb,}j}(\omega)+S_{\mathrm{rp,}j}(\omega)\notag\\
&+S_{\mathrm{th,}j}(\omega)+S_{\mathrm{me,}j}(\omega)].\label{HHH13}
\end{align}
In the coordinate fluctuation spectra, we introduce the effective susceptibility of the $ j $th vibration mode as
\begin{align}
\chi_{j,\mathrm{eff}}(\omega)=\omega_{j}[\Omega_{j,\mathrm{eff}}^{2}(\omega)-\omega^{2}-i\omega\Gamma_{j,\mathrm{eff}}(\omega)]^{-1},\label{HHH14}
\end{align}
where $ \Omega_{j,\mathrm{eff}}(\omega) $ and $ \Gamma_{j,\mathrm{eff}}(\omega) $ are, respectively, the effective mechanical resonance frequency and the effective mechanical decay rate of the $ j $th vibrational mode, defined as
\begin{align}
	& \Gamma_{j,\mathrm{eff}}(\omega)=\gamma_{j}+\gamma_{j,\mathrm{C}}(\omega),\label{HHH15} \\
	& \Omega_{j,\mathrm{eff}}(\omega)=\omega_{j}+\delta\omega_{j}(\omega).\label{HHH16}
\end{align}
The net refrigeration rates $\gamma_{j,\mathrm{C}}$ of the $j$th vibrational modes are
\begin{align}
	\gamma_{1,\mathrm{C}}=&\dfrac{-\left[ G_{1}g_{\mathrm{cd1}}F_{1}(\omega)+2\tilde{\mu}F_{3}(\omega)\right] }{\left[ A_{1}^{2}(\omega)+A_{2}^{2}(\omega)\right] \left[ C_{1}^{2}(\omega)+C_{2}^{2}(\omega)\right] },\notag\\
	\gamma_{2,\mathrm{C}}=&\dfrac{-\left[ G_{2}g_{\mathrm{cd2}}F_{2}(\omega)+2\tilde{\mu}F_{4}(\omega)\right] }{\left[ A_{1}^{2}(\omega)+A_{2}^{2}(\omega)\right] \left[ Y_{1}^{2}(\omega)+Y_{2}^{2}(\omega)\right] },\label{HHH17}
\end{align}
and the mechanical frequency shifts $\delta\omega_{j}(\omega)$ of the $ j $th vibrational mode are caused by the optical spring effect, given by
\begin{align}
	\delta\omega_{1}=&\sqrt{\omega_{1}^{2}+\dfrac{E_{3}(\omega)}{\left[ A_{1}^{2}(\omega)+A_{2}^{2}(\omega)\right] \left[ C_{1}^{2}(\omega)+C_{2}^{2}(\omega)\right] }}-\omega_{1},\notag\\
	\delta\omega_{2}=&\sqrt{\omega_{2}^{2}+\dfrac{T_{3}(\omega)}{\left[ A_{1}^{2}(\omega)+A_{2}^{2}(\omega)\right] \left[ Y_{1}^{2}(\omega)+Y_{2}^{2}(\omega)\right] }}-\omega_{2}.\label{HHH18}
\end{align}
In Eq.~(\ref{HHH13}), we introduced the feedback-induced noise spectrum $ S_{\mathrm{fb,}j}(\omega) $, the radiation-pressure noise spectrum $ S_{\mathrm{rp,}j}(\omega) $, the mechanically-coupling-induced noise spectrum $ S_{\mathrm{me,}j}(\omega) $, and the thermal noise spectrum $ S_{\mathrm{th,}j}(\omega) $ of the $ j $th vibrational mode, which are given by
\begin{align}
S_{\mathrm{th,}j}(\omega)=&\;\dfrac{\gamma_{j}\omega}{\omega_{j}}\coth\label{HHH19}  \beta_{j}   ,\\
S_{\mathrm{me,1}}(\omega)=&\;\dfrac{N_{1}(\omega)}{N_{2}(\omega)}\dfrac{\gamma_{2}\omega}{\omega_{2}}\coth\beta_{2}, \\
S_{\mathrm{me,2}}(\omega)=&\;\dfrac{M_{1}(\omega)}{M_{2}(\omega)}\dfrac{\gamma_{1}\omega}{\omega_{1}}\coth \beta_{1}, \\
S_{\mathrm{fb,1}}(\omega)=&\;\dfrac{\omega^{2}(\kappa^{2}+\omega^{2})\omega_{\mathrm{fb}}^{2}}{4\kappa\vartheta}\dfrac{N_{3}(\omega)}{N_{2}(\omega)},\\
S_{\mathrm{fb,2}}(\omega)=&\;\dfrac{\omega^{2}(\kappa^{2}+\omega^{2})\omega_{\mathrm{fb}}^{2}}{4\kappa\vartheta}\dfrac{M_{3}(\omega)}{M_{2}(\omega)},\\
S_{\mathrm{rp,1}}(\omega)=&\;\kappa\omega_{\mathrm{fb}}^{2}\dfrac{N_{4}(\omega)}{N_{2}(\omega)},\\
S_{\mathrm{rp,2}}(\omega)=&\;\kappa\omega_{\mathrm{fb}}^{2}\dfrac{M_{4}(\omega)}{M_{2}(\omega)},\label{HHH25}
\end{align}
where $ \beta_{j}=\hbar\omega/(2k_{\mathrm{B}}T_{j}) $ and the other used parameters are given in the Appendix~\ref{appendixB}.

\subsection{Giant amplification of both mechanical decay rates and net-cooling rates via the AMC \label{sec4AB}}

We here study how to achieve a giant enhancement in both effective mechanical decay rates $\Gamma_{j,\text{eff}}$ and net-cooling rates $\gamma_{j,\text{C}}$ of the $j$th vibrational mode by introducing the AMC.

In Fig.~\ref{FIG1}(b), we plot the effective mechanical decay rates $\Gamma_{j,\text{eff}}$ as a function of the frequency $\omega$, when the system operates without (i.e., $ \tilde{\mu}=0$, see the solid curves) and with (i.e., $ \tilde{\mu}/\omega_{m}=0.02$, see the dashed curves) the AMC. We find that by introducing the AMC, the effective mechanical decay rates $\Gamma_{j,\text{eff}}$ are largely enhanced at resonance $\omega=\pm\omega_{m}$. Specifically, the effective mechanical decay rates $\Gamma_{\text{eff},j}$ without the AMC (i.e., $\tilde{u}=0$)  are approximately equal to $2\gamma_{j}$ at $\omega=\pm\omega_{m}$, and this means the  cooling of these vibrational modes is inefficient [see the solid curves]. However, when the AMC is switched on (i.e., $\tilde{u}\neq0$), the effective mechanical decay rates $\Gamma_{\text{eff},j}$ at $\omega=\pm\omega_{m}$ can be amplified from $\approx2\gamma_{j}$ to $\gg10^{4}\gamma_{j}$ [see the dashed curves in Fig.~\ref{FIG1}(b)]. This indicates that, by employing the AMC, a giant amplification of the effective mechanical decay rates can be realized, which makes the simultaneous refrigeration of the two vibrational modes feasible.

\begin{figure*}[ht]
	\centering
	\includegraphics[width=17.5cm]{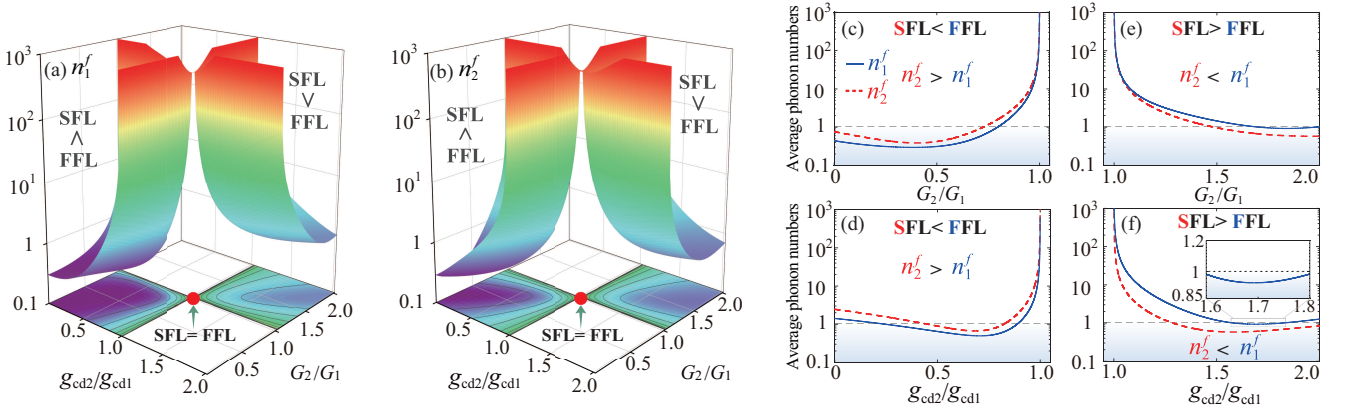}
	\centering
	\caption{ Steady-state  mean thermal phonon numbers (a) $ n_{1}^{f} $ and (b) $ n_{2}^{f} $ in two vibrational modes as functions of the effective optomechanical-coupling ratio $ G_{2}/G_{1} $ and the feedback-gain ratio $ g_{\mathrm{cd2}}/g_{\mathrm{cd1}} $ of the two vibrational modes, when the system operates with the AMC (i.e., $ \tilde{\mu}/\omega_{m} =0.02$). Here the red disks denote the case of SFL$ = $FFL, i.e., $g_{\mathrm{cd},2}=g_{\mathrm{cd},1}$ and $G_{2}=G_{1}$.
 When the strength of the second feedback loop is smaller (larger) than that of the first one [i.e., SFL$ < $FFL (SFL$ > $FFL)], $ n_{1}^{f} $ (blue solid curves) and $ n_{2}^{f}$ (red dashed curves) versus  (c,e) $ G_{2}/G_{1} $  and (d,f) $ g_{\mathrm{cd2}}/g_{\mathrm{cd1}} $. Note that SFL$<$FFL (SFL$>$FFL) describes the parameter conditions: $g_{\mathrm{cd}2}<g_{\mathrm{cd}1}$ and $G_{2}<G_{1}$ ($g_{\mathrm{cd}2}>g_{\mathrm{cd}1}$ and $G_{2}>G_{1}$). Here the parameters used  are : for (c) $ g_{\mathrm{cd1}}=1$ and $g_{\mathrm{cd2}}=0.6 $, (d) $ G_{\mathrm{1}}=0.4\omega_{1}$ and $G_{\mathrm{2}}=0.7G_{1} $, (e) $ g_{\mathrm{cd1}}=1$ and $g_{\mathrm{cd2}}=1.7 $, and (f) $ G_{\mathrm{1}}=0.4\omega_{1}$ and $G_{\mathrm{2}}=1.8G_{1} $. The inset in (f) is a zoomed-in plot of $n^{f}_{1}$ as a function of $ g_{\mathrm{cd2}}/g_{\mathrm{cd1}} $, which clearly shows the dependence
of $n^{f}_{1}$ on $ g_{\mathrm{cd2}}/g_{\mathrm{cd1}} $. Other parameters are the same as those in Fig.\ref{FIG1}.}\label{FIG3}
\end{figure*}

To further illustrate the underlying physics of the multimode refrigeration under the AMC mechanism, we plot the net-refrigeration rate $\gamma_{j,\text{C}}$ of the $j$th vibrational mode versus the AMC strength $\tilde{u}$ at the resonance $\omega=\omega_{m}$, as shown in Fig.~\ref{FIG1}(c). We find that when turning off the AMC (i.e., $\tilde{u}=0$), the net-refrigeration rates are extremely small (i.e., $\gamma_{j,\mathrm{C}}\approx\gamma_{j}$). These results indicate that all the vibrational modes cannot be cooled when the AMC is absent, i.e., $\tilde{u}=0$. However, when we turn on the AMC (i.e., $\tilde{u}\neq0$), the net-refrigeration rates $\gamma_{j,\text{C}}$ are giantly enhanced with the increase of the AMC strength $\tilde{u}$. For example, the net-refrigeration rates $\gamma_{j,\mathrm{C}}$ can be increased from $\gamma_{j,\mathrm{C}}/\gamma_{j}\approx1$ to more than $10^{6}$ in our simulations.

\subsection{Dependence of the multimode optomechanical cooling on the system parameters\label{sec4B}}

In cavity optomechanics, the cold-damping feedback refrigeration of a single vibrational mode can be achieved using the cold-damping effect, which employs a designed feedback force  applied to this vibrational mode, and this leads to the freezing of their thermal fluctuations~\cite{Mancini1998PRL,Genes2008PRA,Steixner2005PRA,Bushev2006PRL,Rossi2017PRL,Rossi2018Nature,Conangla2019PRL,Tebbenjohanns2019PRL,Sommer2019PRL,Guo2019PRL,Sommer2020PRR}. Correspondingly, in principle, the feedback refrigeration of multiple vibrational modes can also be realized based on this feedback-cooling mechanism.

However, in contrast to this anticipation, we find that by using this feedback refrigeration mechanism, a counterintuitive cooling phenomenon emerges, i.e., the multiple vibrational modes cannot be cooled. Physically, the dark mode, which is induced by the coupling of the multiple vibrational modes to a common optical mode, cuts off the thermal-phonon extraction channels. Since the dark mode leads to this counterintuitive uncooling phenomenon, it is natural to ask the question whether one can remove this dark mode to further cool these vibrational modes to their quantum ground states. To this end, the AMC is introduced to our system for removing the dark mode and controlling the refrigeration performance of these vibrational modes.

\begin{figure*}[!ht]
	\centering
	\includegraphics[width=17cm]{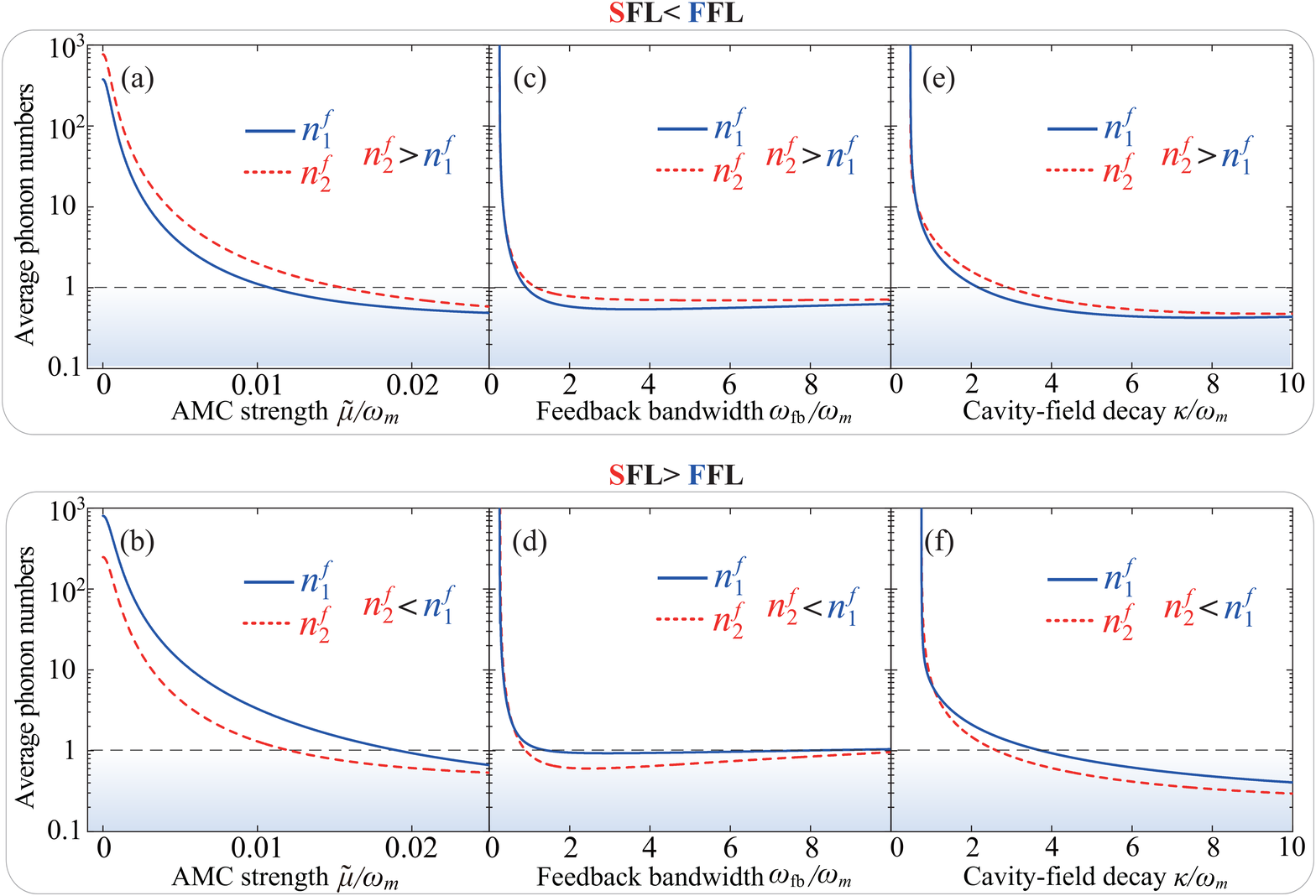}
	\caption{Steady-state mean thermal occupations $ n_{1}^{f} $ (blue solid curves) and $ n_{2}^{f} $ (red dashed curves) versus:  (a, b) the AMC strength $\tilde{u}$,   (c,d) the feedback bandwidth $ \omega_{\mathrm{fb}} $, and (e,f) the cavity-field decay rate $ \kappa$, for the cases when SFL$ < $FFL and SFL$ > $FFL. Other parameters are the same as those in Fig. \ref{FIG3}}\label{FIG4}
\end{figure*}

Specifically, when the AMC is on, we plotted the steady-state mean phonon numbers $n^{f}_{1}$ and $n^{f}_{2}$ versus the optomechanical-coupling-strength ratio $G_{2}/G_{1}$ and the feedback-gain ratio $g_{\mathrm{cd},2}/g_{\mathrm{cd},1}$ of the two vibrational modes, as shown in Figs.~\ref{FIG3}(a) and~\ref{FIG3}(b). It is seen that the two vibrational modes can be efficiently cooled to their quantum ground states ($n^{f}_{1}<1$, $n^{f}_{2}<1$), when the system operates in the regimes for SFL$<$FFL (i.e., $g_{\mathrm{cd},2}<g_{\mathrm{cd},1}$ and $G_{2}<G_{1}$) or SFL$>$FFL (i.e., $g_{\mathrm{cd},2}>g_{\mathrm{cd},1}$ and $G_{2}>G_{1}$). In contrast to the above ground-state refrigeration results, we find that when SFL$=$FFL (i.e., $g_{\mathrm{cd},2}=g_{\mathrm{cd},1}$ and $G_{2}=G_{1}$, see the red disks), the two vibrational modes are not cooled. The physical origin behind this no cooling phenomenon is due to the dark mode, which decouples from the system and prevents the extraction of the phonons. These results mean that the simultaneous ground-state refrigeration of these vibrational modes is achievable owing to the breaking of the system symmetry by introducing the AMC. Breaking the system symmetry leads to the removing of the dark mode.

To further elucidate how the refrigeration performance of the two vibrational modes depends on the parameters of the SFL and FFL, we plot $n^{f}_{1}$ and $n^{f}_{2}$ versus $G_{2}/G_{1}$ [see Figs.~\ref{FIG3}(c) and~\ref{FIG3}(e)] and $g_{\mathrm{cd},2}/g_{\mathrm{cd},1}$ [see Figs.~\ref{FIG3}(d) and~\ref{FIG3}(f)]. We can see from Figs.~\ref{FIG3}(c) and~\ref{FIG3}(d) that when SFL$<$FFL (i.e., $g_{\mathrm{cd},2}<g_{\mathrm{cd},1}$ and $G_{2}<G_{1}$), these mechanical modes can be cooled effectively, and that the refrigeration performance of the first vibrational mode is better than that of the second one. Correspondingly, when SFL$>$FFL (i.e., $g_{\mathrm{cd},2}>g_{\mathrm{cd},1}$ and $G_{2}>G_{1}$), the simultaneous ground-state refrigeration of these vibrational modes can also be realized, and the cooling performance of the second vibrational mode is better than that of the first one. Physically, the strength of the feedback loop directly governs the feedback cooling performance, and this means that the larger is the feedback-loop strength of the resonator, the better is the cooling efficiency of this resonator.

Since the AMC plays a key role in removing the dark mode and achieving the simultaneous refrigeration of the two vibrational modes, the effect of the AMC on the refrigeration performance should be studied in detail.
To this end, we plotted the steady-state mean phonon numbers $n^{f}_{1}$ and $n^{f}_{2}$ as functions of the AMC strength $\tilde{u}$ when SFL$>$FFL and SFL$<$FFL, as shown in Figs.~\ref{FIG4}(a) and~\ref{FIG4}(b). We find that in the absence of the AMC (i.e., $\tilde{u}=0$), neither of the  two vibrational modes can be cooled. In contrast to this, the simultaneous ground-state refrigeration of these vibrational modes is achieved [ i.e., ($n^{f}_{1},n^{f}_{2}<1$)]  by introducing the AMC. This is because by employing the AMC, the dark mode can be completely removed and the refrigeration channels of these vibrational modes can be opened.

In addition, in Figs.~\ref{FIG4}(c) and~\ref{FIG4}(d), we plot $n_{1}^{f}$ and $n_{2}^{f}$ versus the feedback bandwidth $\omega_{\mathrm{fb}}$, when the AMC is on.
We find that the simultaneous ground-sate refrigeration of the two vibrational modes is realized (i.e., $n^{f}_{1},n^{f}_{2}<1$) under the proper parameter conditions, and that the optimal refrigeration of these vibrational modes can be observed for the parameter $\omega_{\mathrm{fb},j}/\omega_{m}>2$. In particular, we demonstrate that, with decreasing the feedback bandwidth, i.e., $\omega_{\mathrm{fb},j}\rightarrow0$, the refrigeration of the two vibrations becomes inefficient. The physical origin behind this is that a smaller feedback bandwidth corresponds to a longer time delay of the feedback loop, and it leads to a lower cooling efficiency for the vibrational modes.

In particular, we find from in Figs.~\ref{FIG4}(c) and~\ref{FIG4}(d) that, when SFL$<$FFL (SFL$>$FFL), the refrigeration performance of the first (second) vibrational mode is better than that of the second (first) one with increasing feedback bandwidth $\omega_{\mathrm{fb},j}$. This asymmetrical cooling is directly induced by the asymmetrical feedback-loop strength, which indicates that the cooling performance is better for a stronger feedback loop.

Furthermore, in Figs.~\ref{FIG4}(e) and~\ref{FIG4}(f), the final average phonon numbers $n^{f}_{1}$ and $n^{f}_{2}$ are plotted as a function of the cavity-field decay rate $\kappa$ when SFL$<$FFL and SFL$>$FFL.
We see that when we decrease the cavity-field decay rate $\kappa$, the refrigeration performance of the two vibrational modes becomes much worse in the resolved-sideband regimes (i.e., $\kappa/\omega_{m}\ll1$). However, by increasing $\kappa$, these vibrational modes are efficiently cooled to their quantum ground states (i.e., $n^{f}_{1},n^{f}_{2}<1$), beyond the resolved-sideband regimes: $\kappa/\omega_{m}>1$. In addition, the optimal refrigeration can be observed for $\kappa/\omega_{m}\geqslant4$. These unresolved-sideband refrigerations are fundamentally different from those in the sideband cooling, for which the optimal cooling is reached only in the resolved-sideband regime~\cite{Wilson-Rae2007PRL,Marquardt2007PRL,Teufel2011Nature}.

\begin{figure}[!htbp]
	\centering
	\includegraphics[width=8cm]{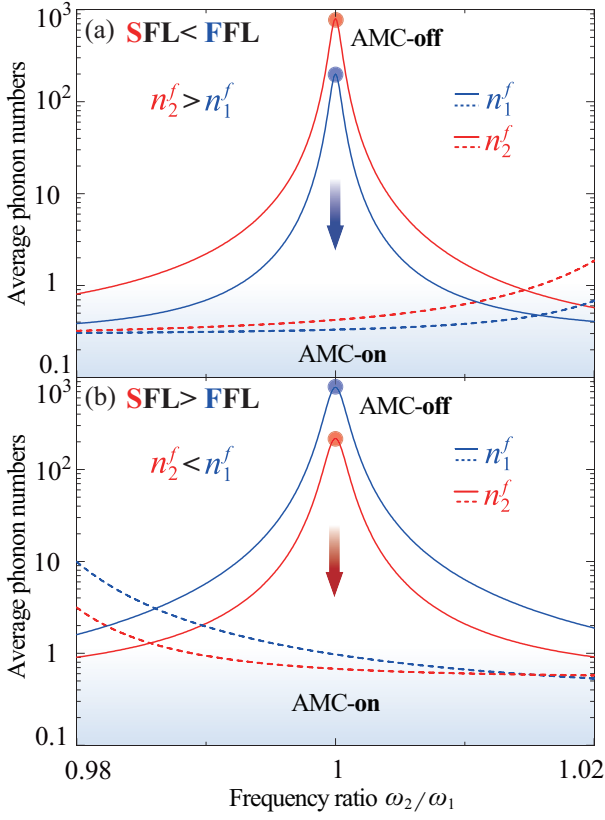}
	\centering
	\caption{Steady-state mean phonon numbers $ n_{1}^{f} $ (blue curves) and $ n_{2}^{f}$ (red curves)  as a function of the frequency ratio $ \omega_{2}/\omega_{1} $ of the two vibrational modes, without ($\tilde{\mu}=0$, solid curves) and with ($\tilde{\mu}=0.02\omega_{1} $, dashed curves) the AMC.  The parameters used for (a) are: $ g_{cd2}=0.5 $ and $ G_{2}=0.5G_{1} $, and  for (b) are: $ g_{cd2}=1.9 $ and $ G_{2}=1.9G_{1} $. Other parameters are set as Fig. \ref{FIG1}.}\label{FIG5}
\end{figure}

We find that, surprisingly, the dark-mode effect can also cause a cooling suppression for the near-degenerate-vibration case. To see the width of the frequency-detuning window associated with this dark-mode effect, in Fig.~\ref{FIG5} we plotted $n_{1}^{f}$ and $n_{2}^{f}$ versus the ratio $\omega_{2}/\omega_{1}$ without  (i.e., $\tilde{u}=0$) and with  (i.e., $\tilde{u}/\omega_{1}=0.02$) the AMC. We find that  without the AMC, the simultaneous ground-state refrigeration of the two vibrational modes is unfeasible in the frequency-detuning range $0.98<\omega_{2}/\omega_{1}<1.02$. However, when the AMC is turned on, the simultaneous ground-state cooling can be realized in the corresponding region. Our findings mean that the AMC mechanism can lead to the simultaneous ground-state refrigeration of both near-degenerate and degenerate vibrational modes. In particular, when SFL$<$FFL (SFL$>$FFL), the cooling efficiency of the first (second) vibrational mode is better than that of the second (first) one. This is because the cooling is governed by the feedback loop, and the cooling performance is better for a larger feedback loop.

\section{Discussion and conclusion\label{sec6}}

\begin{figure}[!htbp]
	\centering
	\includegraphics[width=8cm]{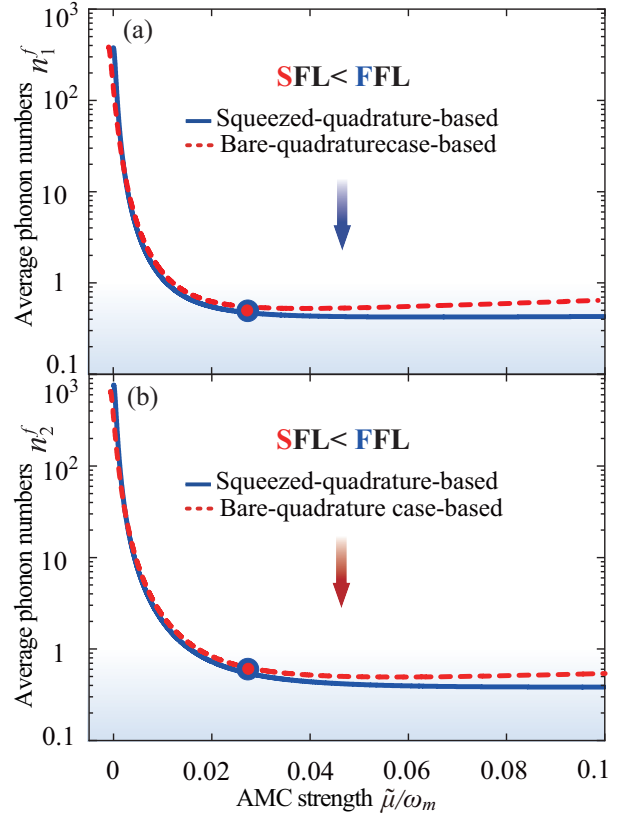}
	\centering
\caption{Steady-state average phonon numbers (a) $n^{f}_{1}$ and (b) $n^{f}_{2}$ are plotted as a function of the AMC $\tilde{\mu}$ for the cooling cases of the squeezed (blue solid curves) and bare (red dashed curves) quadratures, under SFL$<$FFL. Here the bare- and squeezed-quadrature-based cases are, respectively, based on the quadratures in Eqs.~(1) and~(2).}
\label{FigR1}
\end{figure}

Here, we present some discussions to compare the cooling performance based on both bare and squeezed quadratures. It is obvious that the quadratures $q_j$ and $p_j$ in Eq. (2) are squeezed with respect to the bare quadratures, and that our present cooling  pertains to the squeezed quadratures. Due to the fact that the anti-squeezing may have an effect on phonon number, it is worth to further discuss the cooling results in the bare-quadrature-based case. To this end, we plot the final mean phonon numbers $n^{f}_{j}$ as a function of the AMC strength $\tilde{\mu}$ in both bare- and squeezed-quadrature-based cases, as shown in Figs.~\ref{FigR1}(a) and~\ref{FigR1}(b). Here, the blue solid curves show the cooling results in the squeezed-quadrature-based case, while the red dashed curves correspond to that in the bare-quadrature-based case. We find that the two mechanical resonators can be simultaneously cooled to their quantum ground states by introducing the AMC. Physically, the introduced AMC offers an effective strategy to remove the dark mode and, in turn, to rebuild cooling channels for extracting thermal phonons stored in the dark mode. This implies that no actual cooling is observed
for the two mechanical resonators when $\tilde{\mu}=0$. 

Moreover, we find that when $\tilde{\mu}/\omega_{m}\leqslant0.025$, an excellent agreement is observed between the bare-quadrature-based (red dashed curves) and squeezed-quadrature-based (blue solid curves) cooling results, i.e., the cooling performances in both bare- and squeezed-quadrature cases are approximately the same in the region $\tilde{\mu}/\omega_{m}\leqslant0.025$. This can be explained that a weaker AMC strength leads to a smaller squeezing effect with respect to the bare quadratures. In the region $\tilde{\mu}/\omega_{m}>0.025$, the cooling performance of the bare-quadrature-based case becomes worse, while that of the squeezed-quadrature-based case becomes better with increasing $\tilde{\mu}$. 
Physically, by increasing the AMC strength, the quadratures $q_j$ and $p_j$ in Eq. (2) are significantly squeezed with respect to the bare quadratures, and, thus, the cooling performance in the squeezed-quadrature-based case can be improved. These results mean that, when the AMC strength $\tilde{\mu}$ is properly chosen (i.e., $\tilde{\mu}/\omega_{m}\leqslant0.025$), the squeezing effect caused by the AMC has a little effect on cooling performance. However, when $\tilde{\mu}/\omega_{m}>0.025$, the difference of the cooling results in the bare- and squeezed-quadrature-based cases cannot be neglected owing to a significant squeezing effect. Namely, the squeezed-quadrature-based cooling is more efficient than that with the bare quadratures for larger values of the AMC strength, but the differences
between these two cases are negligible if the strength is $\tilde{\mu}/\omega_{m}\leqslant0.025$. Note that in our other simulations, we set $\tilde{\mu}/\omega_{m}=0.02<0.025$,  and this indicates that squeezed-quadrature-based cooling can be used
to implement bare-quadrature-based cooling when $\tilde{\mu}/\omega_{m}\leqslant0.025$.

In summary, we proposed a method to achieve the simultaneous ground-state refrigeration of multiple vibrational modes beyond the resolved-sideband regime, and to realize a giant amplification in the net-refrigeration rates. This is realized by introducing an AMC to break the symmetry of the system and, then, it leads to removing the dark-mode effect. By fully analytical treatments, we showed that when  the AMC is switched on, the amplification of the net-refrigeration rates can be observed for more than six orders of magnitude. Remarkably, we reveal that without  the AMC, the simultaneous ground-state refrigeration of the two vibrational modes is unfeasible; while with the AMC, these vibrational modes can be efficiently cooled to their quantum ground states. Our work could potentially be used for observing quantum mechanical effects and controlling  macroscopic mechanical coherence in the unresolved-sideband regime.

\begin{acknowledgments}
B.-P.H. is supported partly by NNSFC (Grant No.
11974009), the Chengdu technological innovation RD
project (Grant No. 2021-YF05-02416-GX), and the Science Foundation of Sichuan Province of China (Grant
No. 2018JY0180). A.M. is supported by the Polish National Science Centre (NCN) under Maestro Grant No.
DEC-2019/34/A/ST2/00081. F.N. is supported in part by
Nippon Telegraph and Telephone Corporation (NTT) Research, the Japan Science and Technology Agency (JST) [via
the Quantum Leap Flagship Program (Q-LEAP) program, the Moonshot RD Grant No. JPMJMS2061], the Japan Society
for the Promotion of Science (JSPS) [via the Grants-in-Aid
for Scientific Research (KAKENHI) Grant No. JP20H00134],
the Army Research Office (ARO) (Grant No. W911NF18-1-0358), the Asian Office of Aerospace Research and
Development (AOARD) (via Grant No. FA2386-20-1-4069),
and the Foundational Questions Institute Fund (FQXi) via
Grant No. FQXi-IAF19-06.

\end{acknowledgments}

\appendix*

\section{Calculation of the steady-state  mean phonon numbers \label{appendixB}}
In this Appendix, we show the remaining expressions of the parameters used in the Eqs.~(\ref{HHH17})-(\ref{HHH25}). These expressions are defined as:

\begin{widetext}
\begin{align}
A_{1}=&\omega(\kappa+\omega_{\mathrm{fb}}),\notag A_{2}=\kappa\omega_{\mathrm{fb}}-\omega^{2},\notag A_{3}=\omega(\gamma_{2}\omega^{2}-\kappa\Delta_{2}),\notag A_{4}=\omega^{2}(\gamma_{2}\kappa+\Delta_{2}),\notag\\
A_{5}=&\omega\omega_{\mathrm{fb}}(\gamma_{2}\kappa-\omega^{2}),\notag A_{6}=\omega^{2}\omega_{\mathrm{fb}}(\gamma_{2}+\kappa),\notag
A_{7}=G_{2}g_{\mathrm{cd2}}\omega\omega_{2}\omega_{\mathrm{fb}},\notag
A_{8}=\omega\omega_{2}^{2}\omega_{\mathrm{fb}},\notag A_{9}=\kappa\omega_{2}^{2}\omega_{\mathrm{fb}},\notag\\
W_{1}=&\omega(\gamma_{1}\omega^{2}-\kappa\Delta_{1}),\notag
W_{2}=\omega^{2}(\gamma_{1}\kappa+\Delta_{1}),\notag
W_{3}=\omega\omega_{\mathrm{fb}}(\gamma_{1}\kappa-\omega^{2}),\notag
W_{4}=\omega^{2}\omega_{\mathrm{fb}}(\gamma_{1}+\kappa),\notag\\
W_{5}=&G_{1}g_{\mathrm{cd1}}\omega\omega_{1}\omega_{\mathrm{fb}},\notag
W_{6}=\omega\omega_{1}^{2}\omega_{\mathrm{fb}},\notag
W_{7}=\kappa\omega_{1}^{2}\omega_{\mathrm{fb}},\notag
B_{1}=A_{1}A_{3}-A_{2}A_{4},\notag\\
B_{2}=&A_{2}A_{3}+A_{1}A_{4},\notag
B_{3}=A_{1}A_{5}+A_{2}A_{6}+A_{1}A_{7}+A_{1}A_{8}-A_{2}A_{9},\notag\\
B_{4}=&A_{2}A_{5}-A_{1}A_{6}+A_{2}A_{7}+A_{2}A_{8}+A_{1}A_{9},\notag 
C_{1}=B_{1}-B_{3},\notag\\
C_{2}=&B_{2}-B_{4},
\notag D_{1}=2\tilde{\mu}A_{1}-G_{2}g_{\mathrm{cd1}}\omega\omega_{\mathrm{fb}},\notag
D_{2}=2\tilde{\mu}A_{1}-G_{1}g_{\mathrm{cd2}}\omega\omega_{\mathrm{fb}},\notag 
D_{3}=2\tilde{\mu}A_{2},\notag\\
E_{j}=&C_{1}D_{1}D_{j+1}+D_{2}D_{3}C_{3-j}+(-1)^{j}(D_{3}^{2}C_{j}-C_{2}D_{1}D_{4-j}),\notag\\
Y_{j}=&L_{j}-L_{j+2},\notag
L_{j}=A_{j}W_{1}+(-1)^{j}A_{3-j}W_{2},\notag\\
F_{j}=&\omega_{j}\omega_{\mathrm{fb}}x^{2}y^{2}[\gamma_{3-j}^{2}\omega^{2}(\omega^{2}-\kappa\omega_{\mathrm{fb}})\notag\\
&-G_{3-j}g_{\mathrm{cd(3-j)}}\gamma_{3-j}\omega^{2}\omega_{3-j}\omega_{\mathrm{fb}}+\Delta_{3-j}^{2}(\omega^{2}-\kappa\omega_{\mathrm{fb}})],\notag\\
T_{j}=&D_{j+1}D_{1}Y_{1}+(-1)^{j}D_{3}^{2}Y_{j}+(-1)^{j+1}D_{1}Y_{2}D_{4-j}+D_{2}D_{3}Y_{3-j},
\end{align}

\begin{align}
E_{j+2}=&\omega_{1}[E_{j}\omega_{2}(A_{1}^{2}+A_{2}^{2})+(-1)^{j+1}A_{j}G_{1}g_{\mathrm{cd1}}\omega\omega_{\mathrm{fb}}(C_{1}^{2}+C_{2}^{2})],
L_{j+2}=A_{j}(W_{3}+W_{5}+W_{6})+(-1)^{j+1}A_{3-j}(W_{4}-W_{7}),\notag\\
F_{j+2}=&\omega_{1}\omega_{2}x^{2}y^{2}\{ [-(G_{2}g_{\mathrm{cd1}}+G_{1}g_{\mathrm{cd2}})\omega^{2}+2G_{3-j}g_{\mathrm{cd(3-j)}}\tilde{\mu}\omega_{3-j}+(G_{2}g_{\mathrm{cd1}}+G_{1}g_{\mathrm{cd2}})\omega_{3-j}^{2}]\omega_{\mathrm{fb}}(\omega^{2}-\kappa\omega_{\mathrm{fb}})\notag\\
&+\gamma_{3-j}[(G_{2}g_{\mathrm{cd1}}+G_{1}g_{\mathrm{cd2}})\omega^{2}\omega_{\mathrm{fb}}(\kappa+\omega_{\mathrm{fb}})-2\tilde{\mu}xy]\} ,\notag\\
T_{j+2}=&\omega_{2}[T_{j}\omega_{1}(A_{1}^{2}+A_{2}^{2})+(-1)^{j+1}A_{j}G_{2}g_{\mathrm{cd2}}\omega\omega_{\mathrm{fb}}(Y_{1}^{2}+Y_{2}^{2})],
N_{1}=\omega_{2}^{2}f_{+}f_{-},\notag
f_{\pm}=2\tilde{\mu}(\kappa\pm i\omega)(\omega\mp i\omega_{\mathrm{fb}})-G_{2}g_{\mathrm{cd1}}\omega\omega_{\mathrm{fb}},\notag\\
N_{2}=&\omega^{2}x[\gamma_{2}^{2}\omega^{2}+\Delta_{2}^{2}]-2G_{2}g_{\mathrm{cd2}}\omega^{2}\omega_{2}\omega_{\mathrm{fb}}[\omega^{2}\gamma_{2}-\kappa\Delta_{2}]+\{\gamma_{2}^{2}\omega^{2}x+2G_{2}g_{\mathrm{cd2}}\gamma_{2}\kappa\omega^{2}\omega_{2}\notag\\
&+\kappa^{2}\Delta_{2}^{2}
+[\omega^{3}-\omega\omega_{2}(G_{2}g_{\mathrm{cd2}}+\omega_{2})]^{2}\}\omega_{\mathrm{fb}}^{2},\notag\\
N_{3}=& 4g_{\mathrm{cd2}}^{2}\tilde{\mu}^{2}\omega_{2}^{2}+4g_{\mathrm{cd1}}g_{\mathrm{cd2}}\tilde{\mu}\omega_{2}\Delta_{2}
+g_{\mathrm{cd1}}^{2}[\gamma_{2}^{2}\omega^{2}+\Delta_{2}^{2}],
N_{4}=(G_{2}\omega_{2}x_{+}+y_{-})(G_{2}\omega_{2}x_{-}+y_{+}), \notag\\
x_{\pm}=&\omega(G_{2}g_{\mathrm{cd1}}-G_{1}g_{\mathrm{cd2}})/(\kappa\pm i\omega),\notag
y_{\pm}=(\omega\mp i\omega_{\mathrm{fb}})[G_{1}\omega(\omega\mp i\gamma_{2})-2G_{2}\tilde{\mu}\omega_{2}-G_{1}\omega_{2}^{2}]/\omega_{\mathrm{fb}},
M_{1}=\omega_{1}^{2}w_{+}w_{-},\notag\\
M_{2}=&\omega^{2}(\kappa^{2}+\omega^{2})[\gamma_{1}^{2}\omega^{2}+
\Delta_{1}^{2}]-2G_{1}g_{\mathrm{cd1}}\omega^{2}\omega_{1}\omega_{\mathrm{fb}}[\omega^{2}\gamma_{1}-\kappa\Delta_{1}]+\omega_{\mathrm{fb}}^{2}\{\gamma_{1}^{2}\omega^{2}x+2G_{1}g_{\mathrm{cd1}}\gamma_{1}\kappa\omega^{2}\omega_{1}\notag\\
&+\kappa^{2}\Delta_{1}^{2}+[\omega^{3}-\omega\omega_{1}(G_{1}g_{\mathrm{cd1}}+\omega_{1})]^{2}\},~~w_{\pm}=2\tilde{\mu}(\kappa\pm i\omega)(\omega\mp i\omega_{\mathrm{fb}})-G_{1}g_{\mathrm{cd2}}\omega\omega_{\mathrm{fb}},\notag\\
M_{3}=&4g_{\mathrm{cd1}}^{2}\tilde{\mu}^{2}\omega_{1}^{2}+4g_{\mathrm{cd1}}g_{\mathrm{cd2}}\tilde{\mu}\omega_{1}\Delta_{1}+g_{\mathrm{cd2}}^{2}[\gamma_{1}^{2}\omega^{2}+\Delta_{1}^{2}],\notag\\
M_{4}=&(G_{1}\omega_{1}x_{+}+z_{-})(G_{1}\omega_{1}x_{-}+z_{+}). ~~\notag
z_{\pm}=(\omega\pm i\omega_{\mathrm{fb}})[2G_{1}\tilde{\mu}\omega_{1}+G_{2}(\Delta_{1}
\mp i\gamma_{1}\omega)]/\omega_{\mathrm{fb}},
\end{align}
where $j=1,2$, $ \Delta_{j}=\omega_{j}^{2}-\omega^{2} $, $ \Delta_{3-j}=\omega_{3-j}^{2}-\omega^{2} $, $ x=\kappa^{2}+\omega^{2} $, and $ y=\omega_{\mathrm{fb}}^{2}+\omega^{2} $.
\end{widetext}

\end{document}